\documentclass[prb,twocolumn,nofootinbib,aps,preprintnumbers,amsmath,amssymb,superscriptaddress,floatfix]{revtex4-1}
\usepackage{graphicx}
\usepackage{color}
\usepackage{bm}
\usepackage{amsmath}
\usepackage{xspace}
\usepackage[normalem]{ulem}
\usepackage{units}
\usepackage{dcolumn}
\usepackage{paralist}
\usepackage{natmove}
\usepackage{natbib}
\usepackage{hyperref}
\usepackage{epstopdf}

\begin{document}

\title{Phonon spectrum of underdoped $\text{HgBa}_2\text{CuO}_{4+\delta}$ investigated by neutron scattering}

\author{I. Ahmadova}
\affiliation{Department of Physics, University of Colorado at Boulder, Boulder, Colorado 80309, USA}
\author{T. C. Sterling}
\affiliation{Department of Physics, University of Colorado at Boulder, Boulder, Colorado 80309, USA}
\author{A. C. Sokolik}
\affiliation{Department of Physics, University of Colorado at Boulder, Boulder, Colorado 80309, USA}
\author{D.L. Abernathy}
\affiliation{Neutron Scattering Division, Oak Ridge National Laboratory, Oak Ridge, Tennessee 37831, USA}
\author{M. Greven}
\affiliation{School of Physics and Astronomy, University of Minnesota, Minneapolis, Minnesota 55455, USA}
\author{D. Reznik}
\affiliation{Department of Physics, University of Colorado at Boulder, Boulder, Colorado 80309, USA}
\email{}

\date{\today}

\begin{abstract}
The cuprates exhibit a prominent charge-density-wave (CDW) instability with wavevector along [100], i.e., the Cu-O bond direction. Whereas CDW order is most prominent at moderate doping and low temperature, there exists increasing evidence for dynamic charge correlations throughout a large portion of the temperature-doping phase diagram. In particular, signatures of incipient charge order have been observed as phonon softening and/or broadening near the CDW wavevector approximately half-way through the Brillouin zone. Most of this work has focused on moderately-doped cuprates, for which the CDW order is robust, or on optimally-doped samples, for which the superconducting transition temperature ($T_c$) attains its maximum. Here we present a time-of-flight neutron scattering study of phonons in simple-tetragonal $\text{HgBa}_2\text{CuO}_{4+\delta}$ ($T_c = 55$ K) at a low doping level where prior work showed the CDW order to be weak. We employ and showcase a new software-based technique that mines the large number of measured Brillouin zones for useful data in order to improve accuracy and counting statistics. Density-functional theory has not provided an accurate description of phonons in $\text{HgBa}_2\text{CuO}_{4+\delta}$, yet we find the right set of  parameters to qualitatively reproduce the data. The notable exception is a dispersion minimum in the longitudinal Cu-O bond-stretching branch along [100]. This discrepancy suggests that, while CDW order is weak, there exist significant dynamic charge correlations in the optic phonon range at low doping, near the edge of the superconducting dome. 

\end{abstract}

\maketitle

\section{introduction}
Charge-density-wave correlations appear to be universally present in the cuprate phase diagram and have been directly observed in X-ray scattering experiments \cite{blackburn2015x,comin2016resonant,Ghiringhelli821,da2014ubiquitous}. Static CDW order with small modulated charge has been stabilized in a rather narrow doping range, and only at low temperatures. It occurs at a wavevector $\textbf{q}_\text{co}$ close to mid-way between the zone center and the zone boundary, along the direction parallel to the Cu-O bond of the $\text{CuO}_2$ planes. The static CDW order emerges from \textit{dynamic} charge correlations that have been observed in a much larger portion of the phase diagram throughout the superconducting doping range \cite{park2014evidence,chaix2017dispersive,arpaia2019dynamical,yu2019unusual}. It is a distinct possibility that the dynamic charge correlations contribute to the superconducting pairing glue and/or are related to the formation of the pseudogap \cite{yu2019unusual}. These correlations manifest themselves indirectly as a softening and/or broadening of certain phonons near $\text{\bf{q}}_\text{co}$  \cite{reznik2012phonon}. The observed phonon renormalization resembles conventional Kohn anomalies that are characterized by phonon softening at specific wavevectors. However, there is no evidence for this softening from density-functional-theory (DFT) calculations of phonon dispersions in the cuprates \cite{reznik2008photoemission,johnston2010systematic,pintschovius2005electron,reznik2006electron}, whereas such calculations reproduce Kohn anomalies in other compounds rather well \cite{reznik2012phonon,zhang2019pressure,heid2000anomalous}. 

\begin{figure}
\includegraphics[width=1.0\linewidth]{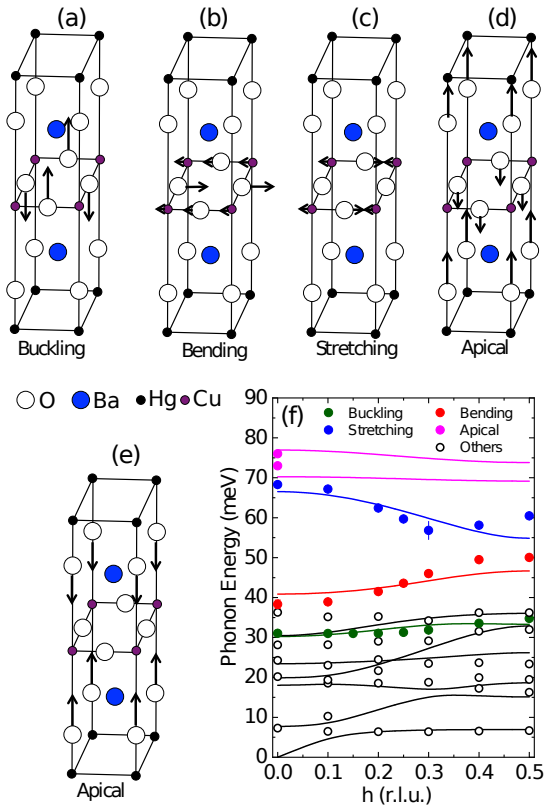}
\caption{Zone-center eigenvectors of select oxygen phonon branches, and phonon dispersions. (a-e) Eigenvectors of select zone-center optic phonons with predominantly oxygen character at the origin of the branches examined in this work. The energies are: (a) 31 meV, (b) 39 meV, (c) 69 meV, (d) 73 meV, and (e) 76 meV. (f) Phonon dispersions of branches with $\Delta_1$ symmetry calculated from DFT (solid lines) and extracted from 10 K neutron scattering data (circles). We were able to observe apical oxygen modes only at the zone center. Error bars are included for all data, but may be smaller than the symbol size. The stretching (blue) and bending (red) dispersions are the same as in Fig. \ref{fig:fig8}c, the apical modes (magenta) are based on the spectra in Fig. \ref{fig:fig9}, and the buckling mode dispersion (green) is the same as in Fig. \ref{fig:fig11}b. All other modes (gray) were determined following the procedure discussed in section IIID. Points above and below 37 meV were obtained based on measurements performed with $E_i = 110 $ meV and 50 meV, respectively.}
\label{fig:fig1}
\end{figure}

The strongest and most common phonon anomaly appears in the Cu-O bond-stretching branch, along [100], which softens and/or broadens close to $\text{\bf{q}}_\text{co}$. This anomaly has been reported for every superconducting cuprate family for which this branch has been measured. However, studies of this anomaly in lightly-doped superconducting cuprates are rare, and it has only been investigated in detail in the ``214" materials $\text{La}_{2-x}\text{Sr}_x\text{CuO}_4$ (LSCO) and $\text{La}_{2-x}\text{Ba}_x\text{CuO}_4$, which exhibit relatively low maximum values of $T_c$ and are prone to charge-spin ``stripe" order.\cite{RevModPhys.75.1201} It was recently shown that charge stripes in the 214 cuprates arise out of universal charge fluctuations at $\text{\bf{q}}_\text{co}$ that are  similar to, or indistinguishable from CDW correlations in other cuprates.\cite{miao2017high}. Interestingly, in LSCO, the strength of the phonon renormalization was found to scale with $T_c$, which points to a possible connection between charge fluctuations and the mechanism of superconductivty \cite{park2014evidence}. 

\begin{figure}
\includegraphics[trim={0 0 0 0},width=1.\linewidth]{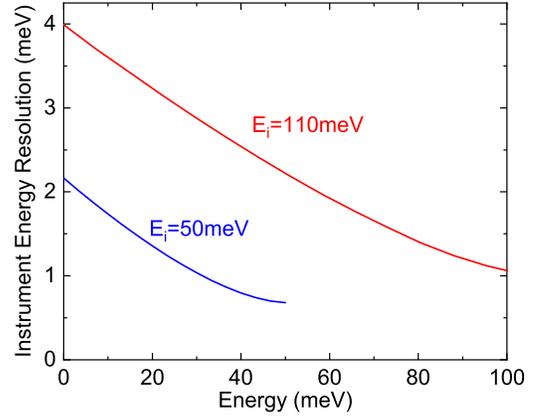}
\caption{Full width at half maximum (FWHM) of the  instrument energy resolution calculated based on Ref. \citenum{Abernathy2012} at the two incident energies used in the experiment.}
\label{fig:ResFunc}
\end{figure}

\begin{figure}
\includegraphics[trim={3.5cm 3cm 4cm 3cm},width=0.8\linewidth]{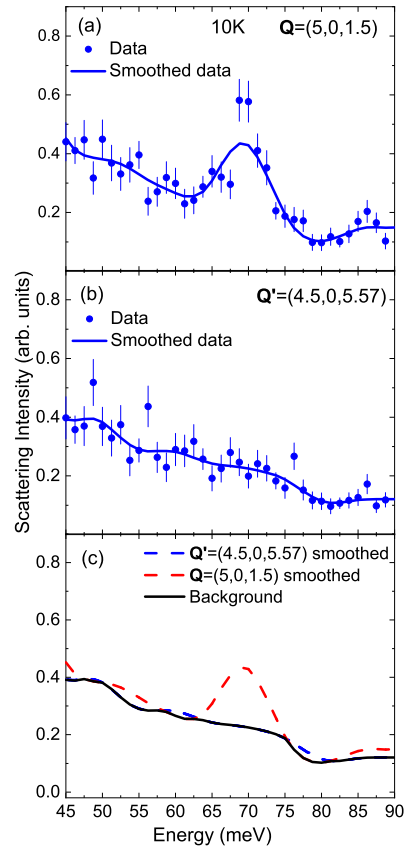}
\caption{Background determination procedure for the bond-stretching phonon at {\bf Q} = (5,0,1.5). Data for (a) {\bf Q} = (5,0,1.5) and (b) {\bf Q'} = (4.5,0,5.57). Solid lines are smoothed data. (c) Black line representing the background is the point-by-point minimum of the two smooth curves in (a,b).}
\label{fig:fig2}
\end{figure}

$\text{HgBa}_2\text{CuO}_{4+\delta}$ (Hg1201) is an exemplary cuprate superconductor, as it features a simple tetragonal crystal structure, minimal point disorder effects , and a maximum $T_c$ of nearly 100 K (at the optimal hole doping level $p \approx 0.16$), the highest value among single-layer cuprates (more than twice the maximum $T_c$ of the 214 cuprates) \cite{eisaki2004effect,barivsic2008demonstrating}. 
This has led to significant insights into the electronic properties of doped CuO$_2$ planes, e.g., through detailed charge-transport \cite{barivsic2013universal,chan2014plane,chan2016single,popvcevic2018percolative,barisic2019evidence,pelc2019resistivity}, magnetic neutron scattering \cite{li2008unusual,li2011magnetic,li2011magnetic2,chan2016commensurate,chan2016hourglass,tang2018orientation}, and X-ray scattering experiments \cite{tabis2017synchrotron,yu2019unusual,d2003phonon,uchiyama2004softening,tabis2014charge}.
Magnetic neutron scattering experiments have revealed no evidence for stripes \cite{chan2016commensurate,chan2016hourglass}, yet there have been no detailed neutron scattering investigations of the phonons in this material. Prior inelastic X-ray scattering work, which focused on optimally-doped Hg1201, reveals a steep downward dispersion in the Cu-O bond-stretching branch, with a minimum midway between the zone center and the zone boundary \cite{d2003phonon,uchiyama2004softening}.   

Here we report results of a comprehensive neutron scattering study of phonons in a very underdoped Hg1201 single crystal with $p \approx0.064$ ($T_c$ = 55 K) \cite{PhysRevB.63.024504,chan2016single}. This doping level is well below $p \approx 0.09$ ($T_c$ = 71 K), where the CDW order was observed to be most prominent \cite{tabis2017synchrotron}. Our emphasis is on the optic modes of predominantly oxygen character illustrated in Fig. 1. 
In order to overcome formidable technical challenges associated with the relatively high neutron absorption of Hg and with the measurement of a relatively small sample, we applied a new software-based technique to analyze time-of-flight (TOF) neutron scattering data. We also performed DFT calculations and found good qualitative agreement with the experiment. One notable exception is the bond-stretching phonon anomaly that appears as a dip in the experimental dispersion that is not predicted by DFT. This feature is weaker than in optimally-doped Hg1201, for which it was previously reported based on inelastic X-ray scattering measurements. \cite{uchiyama2004softening} Nevertheless, the result indicates the presence of considerable dynamic charge correlations near the edge of the superconducting dome at low doping.

\section{methods}
Our new approach to TOF neutron data analysis covers a large volume of reciprocal space including high energy optic phonons, and works best for large (at least 10 g) single-crystalline samples with a relatively small mosaic. Nevertheless, we demonstrate here the feasibility of the technique using a sample formed of about 30 co-aligned crystals, with total mass of about 2 g, and mosaic of appoximately $2^\circ$. Similar Hg1201 samples with higher doping levels were used in prior inelastic magnetic neutron scattering experiments. \cite{chan2016commensurate,chan2016hourglass}
Conventional neutron scattering measurements of detailed phonon dispersions are challenging for such a relatively small sample. An additional complication in the case of Hg1201 is 
the relatively high absorption cross section of Hg. 

We wrote software called Phonon Explorer \cite{reznik} to efficiently mine the large TOF neutron scattering datasets, which vastly improves accuracy and statistics compared to conventional data analysis. This software automates the generation of single-\textbf{Q} energy cuts, background subtraction, fitting, and other tasks, which allowed us to efficiently examine a very large reciprocal space volume of around 100 Brillouin zones. It also allowed us to isolate coherent one-phonon spectra from spurious contamination due to the aluminum sample holder and from sample-intrinsic incoherent scattering. In our notation, an arbitrary reduced wave vector is defined as $\textbf{q} = h \textbf{a}^* +  k \textbf{b}^* +  l \textbf{c}^*= (h,k,l)$ in reciprocal lattice units, with $\textbf{a}^* = \textbf{b}^*$ parallel to the Cu-O bond-direction, and $\textbf{c}^*$ perpendicular to the Cu-O plane, $\textbf{a}^*=2\pi/a$ and $\textbf{c}^* =2\pi/c$, and room-temperature lattice parameters $a=3.88$ \AA \; and $c=9.55$ \AA.
We denote wavevectors in the extended Brillouin zone as {\bf Q} $= (H,K,L)$, with $H, K$ $\parallel$ \{100\} and $L \parallel$ [001].

\subsection{Experimental Details}

The experiment was performed on the ARCS spectrometer at the Spallation Neutron Source with two incident energies: $E_i$ = 50 meV and $E_i$ = 110 meV. Fig. \ref{fig:ResFunc} shows the calculated instrument resolution function.\cite{Abernathy2012} Intrinsic $\bf{Q}$-resolution of the instrument generally varies with $\bf{Q}$, energy and sample shape.\cite{Ehlers2011} In our experiment, the momentum width of the representative Bragg peak at $\bf{Q}$= [4,0,0] was 0.07 $\AA^{-1}$ along \textit{H}, 0.15 $\AA^{-1}$ along \textit{K}, and 0.12 $\AA^{-1}$ along \textit{L}. These values are always smaller than the binning in Q-space used in data analysis, so the effective $\bf{Q}$-resolution is equal to the binning. The measurements were performed at two temperatures: 10 K and 450 K. The sample was mounted with one of the two equivalent $\textbf{a}^*$-axes vertical, and the sample orientation was scanned over 90$^\circ$ about the vertical axis. The scans covered a large part of reciprocal space in the ($H0L$) scattering plane. The \textit{K}-direction was perpendicular to the scattering plane. The coverage along the \textit{K}-direction was about $-1 < K < 1$ for $E_i$ = 50 meV and $-2 < K < 2$ for $E_i$ = 110 meV. 

\begin{figure*}
\includegraphics[trim={1cm 5cm 0 3cm},width=0.8\linewidth]{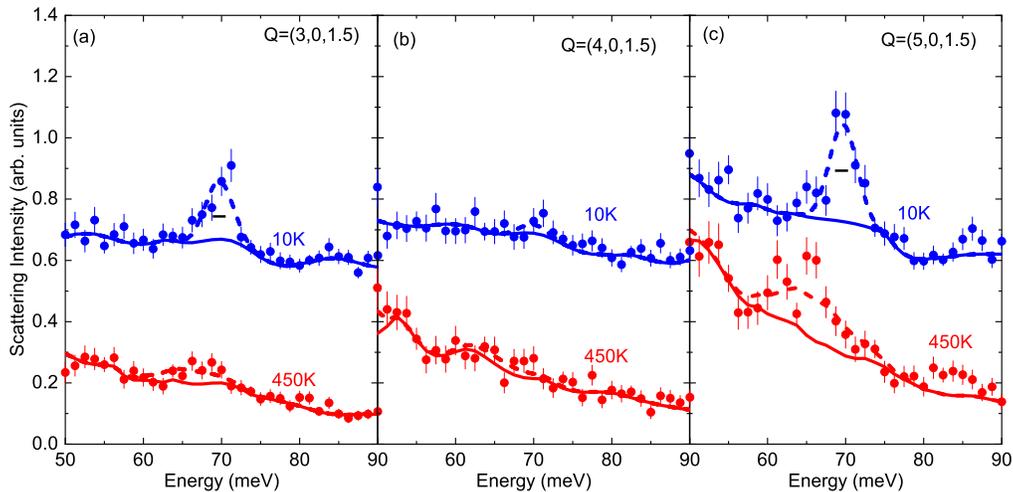}
\caption{Raw data for the zone-center Cu-O bond-stretching phonon. The solid line is the background obtained by the procedure outlined in the text. The dashed line is the sum of the background and the multizone fit to the phonon. The binning was: $\Delta H= \pm 0.1$, $\Delta K= \pm 0.1$, $\Delta L= \pm 3.5$. The blue curves are shifted up by 0.5 for clarity. Short black horisontal line represents FWHM of the calculated instrument resolution.}
\label{fig:fig3}
\end{figure*}

\subsection{Data Analysis}
The data analysis procedure utilizes the Phonon Explorer software.\cite{reznik} that leverages the entire data set to overcome the relatively poor counting statistics in any single constant-\textbf{Q} energy cut. For example, to look at the zone-center bond-stretching phonon, we made constant-\textbf{Q} cuts at all the zone-center wavevectors with the momentum transfer primarily along [100]: $\bf{Q}=$ (3,0,0), (4,0,0), (5,0,0), (4 1,0) and (4,-1,0). We integrated along [001] from $L =-2$ to $L=5$ (there exist no data for $L <-2$). The integration range along [001] was chosen such that there was no observable $L$-dependence in the data.

An accurate background determination was necessary in order to obtain the best results. Phonon Explorer has a built-in user-controlled background determination and subtraction framework, wherein the background is effectively read out from regions of (\textbf{Q},$\omega$) in which one-phonon scattering intensity is negligible. 

In the case of the bond-stretching phonon branch, we determined the background as follows:

\medskip
    At each \textbf{Q}:
    
    \begin{enumerate}
        \item Pick another wavevector, \textbf{Q'}, as close to \textbf{Q} as possible, such that $|$\textbf{Q'}$|=|$\textbf{Q}$|$ with no significant one-phonon scattering in the energy-range of the phonon of interest; 
        \item Generate energy cuts at both \textbf{Q} and \textbf{Q’} and smooth both; 
        \item The background is the point-by-point minimum of the two curves.
    \end{enumerate}

Restriction to the same wavevector magnitudes in step 1 ensured the removal of powder contributions to the scattering. 

\begin{figure}
\includegraphics[trim={1cm 8cm 1cm 4cm,clip},width=1\linewidth]{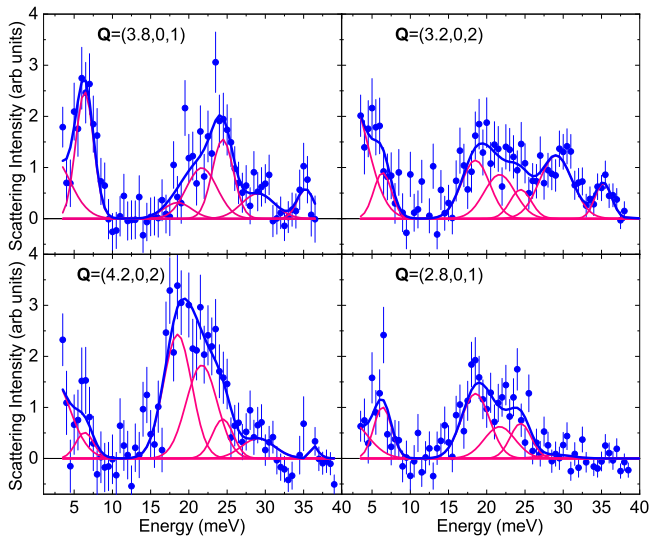}
\caption{Representative datasets used in the multizone fit to the low-energy raw neutron scattering data at the reduced wavevector {\bf q} = (0.2,0,0). The fit was based on 11 datasets at wavevectors that correspond to the same reduced wavevector. Peak positions were determined to be 6.4, 18.5, 21.7, 24.4, 29.1, and 35.2 meV.  The background was assumed to be linear in energy and to run through the minima of the data. The data were taken at 10 K with $E_i$ = 50 meV. The binning was: $\Delta H=0.05$, $\Delta K=0.1$, $\Delta L=0.5$. The points determined from the fitted peak positions correspond to some of the gray circles in Fig. \ref{fig:fig1}f.}
\label{fig:fig4}
\end{figure}

This procedure is illustrated in Fig. \ref{fig:fig2}. The energy cut at {\bf Q} $ = (5,0,1.5)$ for which we wish to determine the background is shown in Fig. \ref{fig:fig2}a. The peak at 69 meV originates from the bond-stretching phonon. It is known that this branch disperses downward toward the zone boundary,\cite{uchiyama2004softening} and therefore we do not expect any phonons near this energy at $H=4.5$. In order to keep the magnitude of the wavevector the same, we increase the magnitude of $L$ to 5.57, so \textbf{Q'} $=(4.5,0,5.57)$. The cut at \textbf{Q'} is shown in Fig. \ref{fig:fig2}b. The solid lines in Figs. \ref{fig:fig2}a,b represents the result of our smoothing procedure, which involved an automatic fit to a set of closely-spaced Gaussian peaks. Step 3 is illustrated in Fig. \ref{fig:fig2}c. 

Raw data around 69 meV at three zone-center wavevectors together with the background determined by our procedure are shown in Fig. \ref{fig:fig3}. This procedure can be generalized to more than one \textbf{Q'} wavevector, if necessary. 

This algorithm does not enable the determination of the background for the data below approximately 30 meV, because some appreciable one-phonon intensity appears at all wavevectors. In this case, we assumed a linear background that goes through the minima of the cut (Fig. \ref{fig:fig4}).

\begin{figure}
\includegraphics[trim={7cm 5cm 7cm 2cm,clip},width=0.3\linewidth]{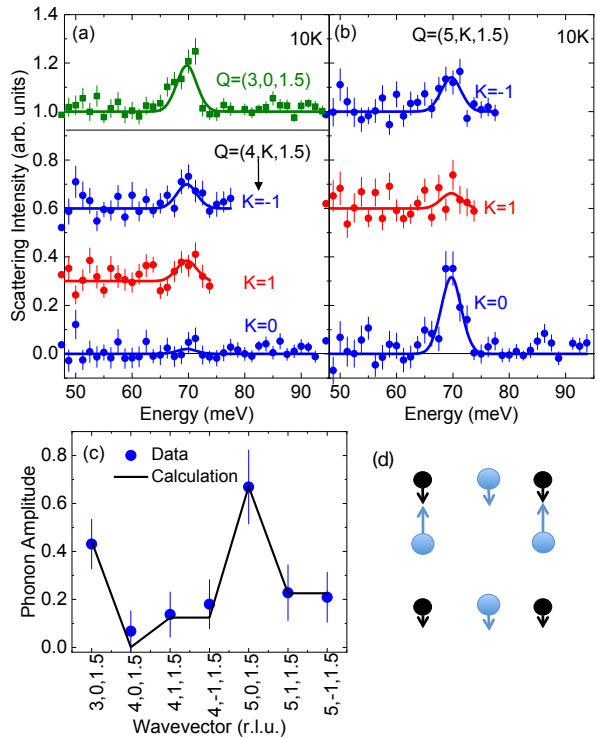}
\caption{Zone-center bond-stretching phonon. (a,b) Solid line is the result of a multizone fit (see text) to the background-subtracted data (symbols). (c) Integrated intensity of the phonon at different wavevectors compared with the prediction based on the eigenvector shown in (d). (d) Eigenvector of the bond-stretching phonon. Small black circles represent Cu ions, large blue circles represent O. Arrows represent the phonon eigenvector drawn to scale. The eigenvector was extracted from phonon intensities in (a,b)}
\label{fig:fig5}
\end{figure}

Background-subtracted data were analyzed using multizone fitting \cite{parshall2014phonon}, whereby peak positions and linewidths for phonons at a given reduced wavevector were fixed to be the same in different Brillouin zones according to Bloch's theorem, but amplitudes were allowed to vary independently. Results of such fits (on top of the estimated background) are shown as dashed lines in Fig. \ref{fig:fig3}, and as solid lines in Fig. \ref{fig:fig4}. Multizone fitting was found to greatly improve the precision of the peak position and linewidth determination, as it effectively combines counting statistics from different Brillouin zones. It also allowed the extraction of phonon eigenvectors from phonon intensities if sufficiently many Brillouin zones were included (Fig. \ref{fig:fig5}). The fit function was a sum of Gaussian peaks and did not explicitly include instrument resolution. The latter would have required a significant upgrade of the software, since the resolution function of TOF instruments has a complex and asymmetric functional form. This upgrade is planned for the future and outside the scope of the present work.

Using data from more than one Brillouin zone automatically differentiates phonon peaks in the constant-$\bf{Q}$ scans from spurious peaks such as multiple scattering peaks, since the former obey Bloch's theorem and the latter do not. Note that we do not attempt to separate multiple-scattering and other contributions to the background from the data. 

\subsection{DFT Calculations}
We found that DFT calculations for Hg1201 give different phonon dispersions depending on the approximations that were made. Here we discuss the calculation that gave the best agreement with the data, as illustrated in Fig. 1f.

We calculated ab-initio lattice dynamical properties of Hg1201 in the harmonic approximation using the frozen-phonon technique as implemented in the PHONOPY package \cite{togo2015first}. For the force calculations, we employed DFT within the local density approximation \cite{goedecker1996separable}. A combination of atom-centered Gaussian functions and plane waves was used to represent the electronic wave functions and density, respectively, as implemented in the Quickstep \cite{vandevondele2005quickstep} algorithm in the CP2K code \cite{hutter2014cp2k}. The MOLOPT-DZVP-SR-GTH \cite{vandevondele2007gaussian} Gaussian basis-set type was used for all atoms along with a plane-wave cutoff of 1000 Ry for the finest integration grid and a $1\times10^{-10}$ Ry convergence requirement on the total energy. The O-$2s$, O-$2p$, Hg-$5d$, Hg-$6s$, Cu-$3d$, Cu-$4s$, Ba-$5p$ and Ba-$6s$ electrons were included as valence states and norm-conserving pseudopotentials of the GTH type \cite{hartwigsen1998relativistic} were used to represent all others. 

Before force constants were calculated, we optimized the atomic positions until the maximum force on any atom was less than $1\times10^{-8}$ Hartree/Bohr while keeping the tetragonal unit cell fixed at the experimental lattice parameters. During optimization, Brillouin-zone integrations were performed using a 12x12x10 Monkhorst-Pack $k$-point grid and Fermi-Dirac smearing with a 300 K effective temperature. In order to compute the force constants, 2x2x2 supercells were constructed from the optimized unit cell and a reduced 6x6x5 $k$-point grid was used for Brillouin zone integrations. The supercell dimensions were chosen so that contributions from periodic images were negligible; we determined this by observing that the forces on the atoms furthest from the displaced site are more than an order of magnitude smaller than the forces on the closest atoms within the supercell. A total of seven supercells with symmetry inequivalent displacements of 0.02 Bohr were used to construct the force constant matrix. The phonon dispersions and eigenvectors were calculated from the force constants using PHONOPY. 

In order to determine which DFT phonon branches have non-negligible intensity for the selected constant \textbf{Q} slices analyzed with the Phonon-Explorer software, we computed the theoretical neutron scattering intensities using the SNAXS software package\cite{parshall_2016}. For the calculation of the theoretical neutron scattering intensities, we used an incident neutron energy $E_i$ of 50 meV. Comparisons between longitudinal DFT phonon dispersions (solid lines) and longitudinal optic branches determined from the neutron data along {\bf q} $=(h,0,0)$ are presented in Fig. \ref{fig:fig1}f.

\section{results}

The main scientific goal was to determine whether a bond-stretching phonon anomaly is present in strongly underdoped Hg1201 and to search for an effect in the Cu-O plane bond-buckling branch. Therefore, we primarily focused on optic phonons of predominantly oxygen character, whose zone-center eigenvectors \cite{stachiotti1995lattice} are illustrated in Fig. \ref{fig:fig1}. Specifically, we carefully measured the dispersion of the bond-stretching branch (Fig. \ref{fig:fig1}c), distinguishing it from the Cu-O plane bond-bending branch (Fig. \ref{fig:fig1}b), idendified the apical oxygen phonons in order to learn about their possible anticrossing with the bond-stretching branch (Fig. \ref{fig:fig1}d,e), and measured the bond-buckling branch (Fig. \ref{fig:fig1}a) with high precision at both 10 K and 450 K. We also measured dispersions of the longitudinal phonons below 45 meV along [100] to ensure that the DFT calculations gave the correct results. Our measurements also included a search for possible signatures of incipient CDW order in the lowest optic branch around 6 meV. Because the sample was mounted in the ($H0L$) scattering plane, the natural reciprocal space direction for the phonon dispersion in the $ab$-plane is along [100]. Access to the [110] direction was limited to the first and parts of the second $ab$-plane Brillouin zones, which precluded measurements of LO and LA phonons, but allowed investigation of some TO modes in this direction. 

\begin{figure}
\includegraphics[trim={0cm 1cm 0cm 1.5cm},width=0.85\linewidth]{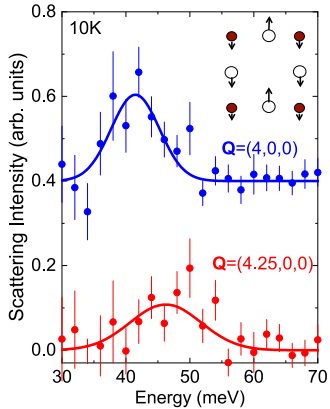}
\caption{Representative background-subtracted spectra of the bond-bending phonon branch. The data were analyzed using multizone fitting, which included other wavevectors at the same reduced \textbf{q}: for $h=0$, {\bf Q} $= (4,0,0), (4,1,0)$ and (4,-1,0); for $h=0.25$, {\bf Q} $=(4.25,0,0), (4.25,1,0)$ and (4.25,-1,0). The inset represents a qualitative schematic of the zone-center phonon eigenvector consistent with strongest intensity at \textbf{Q}=(4,0,0). The data were of insufficient quality for a quantitative determination as was done in Fig. \ref{fig:fig5}.
}
\label{fig:fig6}
\end{figure}

\subsection{Bond-Stretching and Bond-Bending Phonons}

We start our discussion with the bond-stretching phonon branch, which is the primary focus of this work. Background-subtracted zone-center spectra analyzed using the multizone fitting procedure are shown in Fig. \ref{fig:fig5}. Even though the individual scans are rather noisy, the multizone fit gives a very accurate phonon energy of 68.3 $\pm$ 0.2 meV.

Representative background-subtracted data for the bond-bending branch at the zone center and half-way to the zone boundary are shown in Fig. \ref{fig:fig6}. The multizone fit includes additional energy cuts (not shown) at \textbf{Q} $=(4,1,0)$ and (4,-1,0). 

\begin{figure}
\includegraphics[trim={0cm 2cm 0cm 1cm},width=1.5\linewidth]{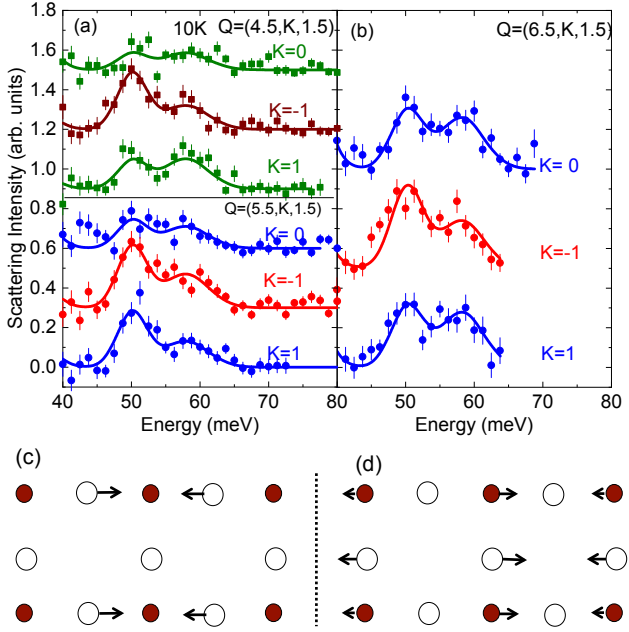}
\caption{X-point zone-boundary longitudinal phonons above 40 meV. (a,b) Spectra were fit
to two peaks using multizone analysis. The binning was $\Delta H = \pm 0.12$, $\Delta K = \pm 0.15$, and $\Delta L= \pm 3.5$.
(c,d) Eigenvectors of these phonons. The intensity variation with $K$ is consistent with (d) the
bending phonon at the lower energy, 50meV, and (c) the stretching (half-breathing) phonon at the
higher energy, 58.5meV.}
\label{fig:fig7}
\end{figure}

\begin{figure}
\includegraphics[trim={1cm 1cm 0cm 2cm},width=1.0\linewidth]{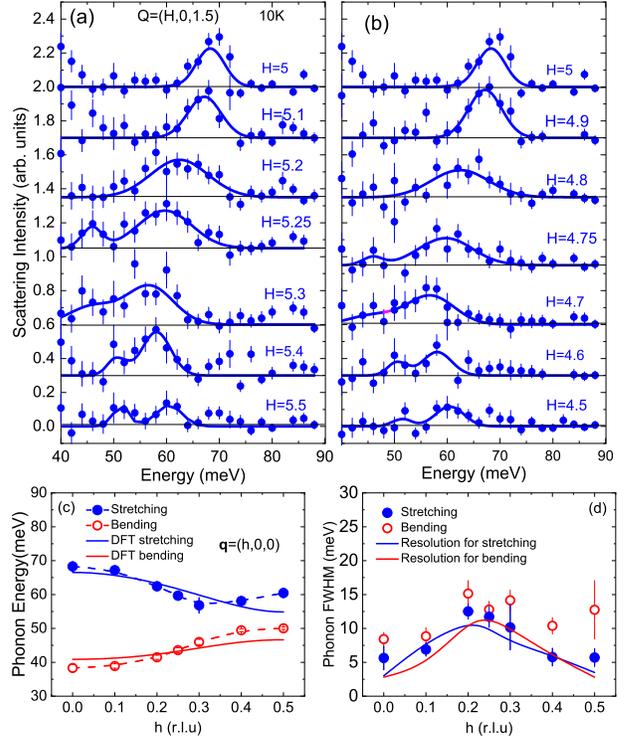}
\caption{Cu-O bond-stretching and bond-bending longitudinal phonon dispersion along [100]. (a,b) Background-subtracted data at wavevectors with large bond-stretching phonon structure factor and small bond-bending structure factor. Binning: $\Delta H= \pm 0.07$, $\Delta K= \pm 0.08$ and $\Delta L= \pm 3.5$. (c) Dispersions of the bond-stretching (blue) and bond-bending phonon branches (red). For the bond-stretching phonon peaks the multizone fit was made to the data at each $q$ in the two zones shown in (a,b). The bond-bending (lower) branch peak positions were fixed to the values obtained based on wavevectors adjacent to (4,0,0), (4,1,0) and (4,-1,0), with representative scans shown in Fig. \ref{fig:fig6}. The width of the bond-bending branch was a free parameter, because the binning was different from Fig. \ref{fig:fig6}. (d) Raw-data peak widths (FWHM) extracted from the same fits as in (c). The effective resolution width increases when the dispersion is larger inside the binned \textbf{Q}-space volume.}
\label{fig:fig8}
\end{figure}

Zone-boundary spectra containing both the bending (50 meV) and stretching (59 meV) phonons are shown in Fig. \ref{fig:fig7}a,b. Note that the data in Fig. \ref{fig:fig8} suggest a somewhat higher energy of 60.3 meV compared with 59 meV in Fig \ref{fig:fig7}, which is due to a tighter binning in momentum space in Fig. \ref{fig:fig8}. This discrepancy is consistent with an upward dispersion of this branch on toward the zone boundary. 

We can distinguish the two branches by their amplitude variation from zone to zone. The bond-stretching branch involves only the motion of plane oxygen atoms, as shown in Fig. \ref{fig:fig7}c, and therefore the scattering intensity from this phonon should be independent of $K$. On the other hand, the bending phonon involves Cu and the other two plane oxygens, as illustrated in Fig. \ref{fig:fig7}d. This eigenvector gives constructive interference between scattering from O and Cu for $K=\pm 1$, and destructive interference for $K=0$. Since the intensity enhancement at $K=1$ is seen only in the 50 meV mode, we assign it to the bending branch, and the 59 meV mode to the stretching branch. Based on this determination, we conclude that the bending and stretching branches do not cross. Eigenvectors extracted from our DFT results support this conclusion.

The dispersion of the bond-stretching branch is shown in Fig. \ref{fig:fig8}. The phonon starts out with a zone-center energy of 68.3 meV, and then disperses downward along [100] toward the zone boundary, with a minimum around $h=0.3$. The phonon peak broadens away from the zone center, but this broadening is consistent with the dispersion-broadened resolution (Fig. \ref{fig:fig8}d). For example, at $h=0.2$, due to the relatively steep dispersion, the phonon energy at one end of the binning range is about 7 meV higher than at the other end. This causes an effective increase of the width compared to the zone center or zone boundary, where there is little dispersion. Thus we do not observe any measurable enhanced intrinsic broadening of the bond-stretching (or bond-bending) phonons around $\textbf{q}_{\text{co}}$. DFT calculations predict a smooth downward dispersion of the bond-stretching phonon branch, and an upward dispersion of the bond-bending branch (Fig. 8c). These calculated dispersions qualitatively reproduce the data, but they do not show the dip near $h=0.3$ that the bond-stretching branch exhibits. This experimentally-determined dispersion minimum in the bond-stretching branch has been interpreted as a signature of anomalous phonon softening. \cite{reznik2012phonon}

\subsection{Apical Oxygen Phonons}

It is important to know whether or not transverse optic branches along [100] that originate from the phonons characterized by apical oxygen displacements along the c-axis cross the bond-stretching branch. There should be two branches with apical oxygen bond-stretching character: one that originates from the Raman-active zone-center mode, where the apical oxygens move in opposite directions, and another one that originates from the IR-active zone-center mode, where the apical oxygens move in the same direction against the Cu/Hg. We found these branches at 73 meV and 76 meV. The background was obtained with \textbf{Q'}$=(H,0,10)$, because the intensity of both phonons vanishes at $L=10$. 

\begin{figure}
\includegraphics[trim={1cm 10cm 1cm 2.5cm},width=1.3\linewidth]{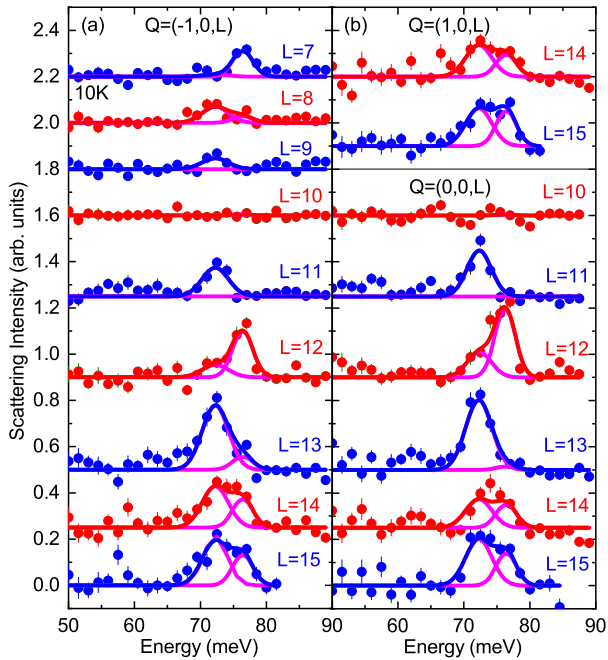}
\caption{Background-subtracted apical-oxygen bond-stretching phonon spectra. Solid lines represent multizone fit results based on all the data shown in the figure. The data are vertically offset for clarity. Binning: $\Delta H= \pm 0.2$, $\Delta K = \pm 0.2$ and $\Delta L = \pm 0.3$.}
\label{fig:fig9}
\end{figure}

In contrast with the in-plane vibrations, we found strong L-dependence of these modes. Cuts with different $L$-values show very different results and, it was therefore not possible to bin the data with large $L$. In order to improve statistics, we used a fairly large binning along $H$ and $K$ after verifying that the phonon energies do not depend on $H$ or $K$. The data around the zone center at many different $L$-values are shown in Fig. \ref{fig:fig9}. Multizone fitting clearly distinguishes between the two peaks, although both of them merge at some $L$-values, whereas at other $L$-values only one peak is clearly visible. 

The rapid zone-to-zone intensity variation along [001] is not consistent with the simple eigenvectors proposed in Ref. \citenum{stachiotti1995lattice}. The most likely scenario is that the branches start to mix just a small distance from the zone center, within our binning range, so that the neutron data do not reflect the zone-center eigenvectors. We were not able to follow these branches further along [100] due to poor statistics, but it is safe to assume that they should be rather dispersionless due to their transverse character and well above the bond-stretching branch. Interactions between the bond-stretching branch and the apical-oxygen branches ought to be negligible, because of this difference in energy. Our DFT results indeed reveal a flat dispersion of the apical modes and show no sign of mixing between the apical and the bond-streching branches (Fig. 1f). 

\subsection{Bond-buckling Phonons}

In addition to the bond-stretching phonon anomaly, a similar, but weaker effect has been reported for an optic phonon (buckling) branch of $\text{YBa}_2\text{Cu}_3\text{O}_{6+\delta}$ (YBCO) that is characterized by planar oxygen vibration in the direction perpendicular to the $\text{CuO}_2$ planes, \cite{reznik1994high,raichle2011highly}. 

In Hg1201, the equivalent zone-center mode is the silent $\text{B}_{2\text{u}}$ mode that cannot be seen in Raman or IR spectra. We also looked at the branch that originates from this mode (Fig. \ref{fig:fig1}a). In addition, this phonon branch is interesting due to its unusual structure factor: the zone-boundary buckling phonon at the two-dimensional reduced wavevector $(\pi/a,\pi/a)$ can be misidentified as a magnetic excitation \cite{fong1995phonon}. We carefully measured its dispersion to see if the anomalous behavior seen in YBCO is also present in Hg1201.
We note that this phonon was identified in previous neutron scattering work \cite{chan2016commensurate,chan2016hourglass} that focused on the magnetic excitations in underdoped Hg1201 near the antiferromagnetic wavevector $(\pi/a,\pi/a)$ and systematically excluded from the data analysis, but not studied in detail.

\begin{figure}
\includegraphics[trim={1cm 2cm 0cm 1cm},width=1.0\linewidth]{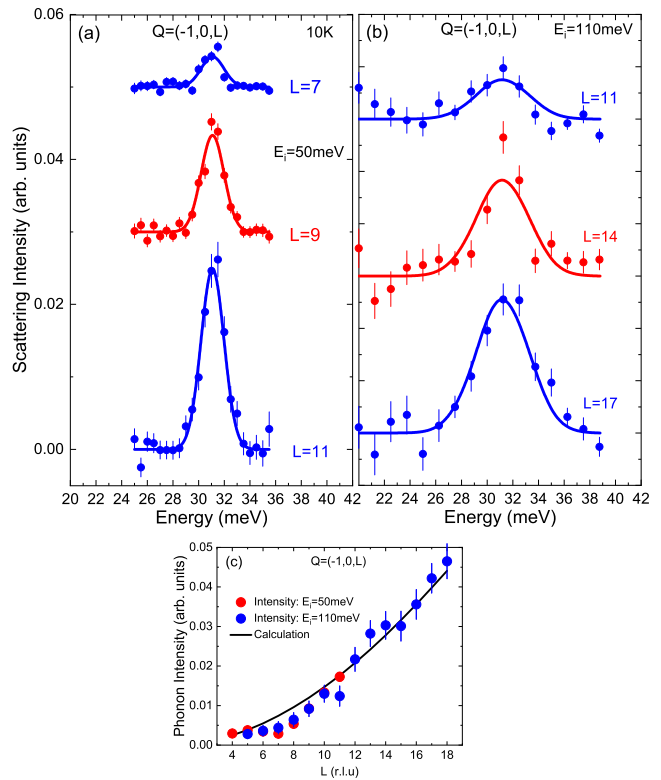}
\caption{$\text{B}_{1\text{u}}$ silent mode at the zone center. The phonon was measured with incident energies of (a) 50 meV and (b) 110 meV. Binning: $\Delta H= \pm 0.1$, $\Delta K= \pm 0.1$ and $\Delta L= \pm 0.5$. (c) Phonon intensity as a function of $L$ compared with the calculation based on the phonon eigenvector and assuming no disorder.}
\label{fig:fig10}
\end{figure}

This mode can be (and has been, in the case of YBCO) mistaken for a magnetic excitation because its structure factor is strong at $(\pi/a,\pi/a)$ and vanishes at $h=k=0$ and $h=k=1$. A momentum scan at the phonon energy from the two-dimensional zone center to $(H,K)=(1,1)$ will therefore give rise to a peak at $(\pi/a,\pi/a)$. The structure factor for this phonon at the zone center is largest for {\bf Q} $=(\pm1,0,L)$ or {\bf Q} $=(0,\pm1,L)$.\cite{fong1995phonon} We found that this phonon appears near 31 meV, as shown in Fig. \ref{fig:fig10}. The background is easy to determine by going to a nearby wavevector of {\bf Q'} $=(0,0,L)$, so the phonon intensity extracted from multizone fits is quite reliable. We used the intensity of this phonon to find out if there is any disorder of the plane oxygens along [001]. Disorder influences the functional form of the intensity of the phonon as a function of $L$ (Fig. \ref{fig:fig10}c) via the Debye-Waller factor. We found that the intensity as a function of $L$ is consistent with negligible disorder effects (the fit gives $6\times10^{-4}\pm6\times10^{-5}~\text{\r{A}}$). 

\begin{figure}
\includegraphics[trim={1cm 8cm 0cm 1cm},width=1.0\linewidth]{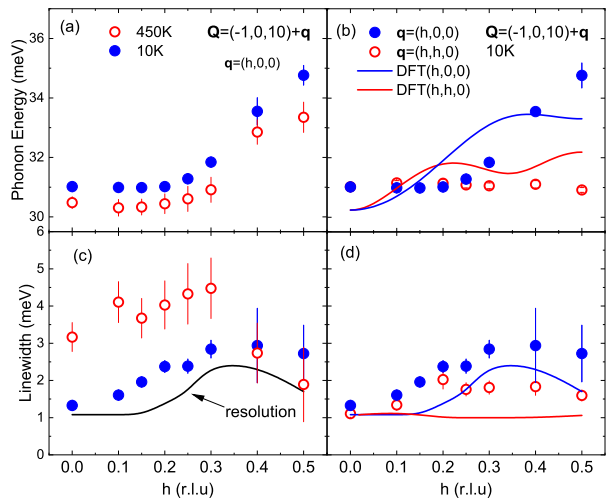}
\caption{ Dispersion and linewidth of the bond-buckling phonons along [100] and [110] at 10 K (b,d) and also at 450K for [100] (a,c). Binning: $\Delta H=0.05$, $\Delta K=0.1$, $\Delta L=2.5$ at 10 K, and $\Delta H=0.1$, $\Delta K=0.1$, and $\Delta L=2.5$ at 450 K.
Red/blue lines in (d) represent experimental resolution for {\bf q} $=(h,h,0)/(h,0,0)$ respectively. Zone center data are shown in Fig. 11.}
\label{fig:fig11}
\end{figure}

We observe a significant dispersion anisotropy between [100] and [110] (Fig. \ref{fig:fig11}b). The dispersion along [100] turns up around $h=0.25$ at both 10 K and 450 K whereas the dispersion along [110] is flat. The apparent phonon broadening in the range $h=0.25-0.5$ can be attributed to the dispersion-enlarged effective energy resolution and, unlike for YBCO, is therefore not intrinsic.

\subsection{Low-Lying Branches}

Recently, attention has focused on the lowest optic branch, because in YBCO it shows dispersion dips and/or linewidth enhancement around $h=0.3$, where the CDW has been observed.\cite{blackburn2015x,comin2016resonant,Ghiringhelli821} Ref. \citenum{d2003phonon} found that low-energy phonon dispersions in optimally-doped Hg1201 were consistent with a Born-van-Karman model calculation and did not report any anomalous behavior.  
We carefully looked at this branch and found no dispersion dips, in agreement with previous work for optimally-doped Hg1201 \cite{d2003phonon}. Its linewidth is resolution-limited, and the spectra at 10 K and 450 K are indistinguishable, apart from the Bose factor. 

We also extracted phonon dispersions for branches with $\Delta_1$ symmetry (Fig. 1f) ($\Delta_1$ symmetry refers to phonon eigenvectors that are symmetric with respect to reflections about the \textit{ab}-plane and about the \textit{ac}-plane) to see how well the DFT calculation matches the data. Peak positions were obtained using multizone fitting \cite{parshall2014phonon} (Fig. \ref{fig:fig4}). We find good qualitative agreement between the experimental data and the DFT results for all the branches considered.

\section{Discussion and Conclusions}

In the current work, we used the Phonon Explorer software to unambiguously identify dispersions of most $\Delta_1$ phonon branches along [100] based on inelastic neutron scattering data taken on a relatively small sample of Hg1201. We found that branch-crossing interplay is not responsible for the bond-stretching phonon dispersion dip because the apical oxygen branches are at higher energy than the bond-bending and bond-stretching branches. As a result, they should not interact with the bond-stretching branch, which disperses downward from the zone-center energy of 68.3 meV. The longitudinal optic (LO) Cu-O bond-bending branch starting from the zone center phonon illustrated in Fig. \ref{fig:fig1}f disperses upward from 41 meV to 50 meV at the zone boundary. Analysis of phonon intensities at the zone boundary showed that the  bond-stretching and bond-bending branches retain their pure character and do not mix. This observation is consistent with the claim that the strong coupling of the bond-stretching branch with dynamic charge fluctuations generated via strong nonlocal electron phonon interaction is the origin of its generic softening in the cuprates \cite{PhysRevB.73.224502,bauer2010phonon,Bauer_2010_2}.

We emphasize that none of our DFT calculations reproduced the experimentally-observed minimum in the bond-stretching phonon dispersion. This is a well-known property of all phonon anomalies away from the zone center in the cuprates, and likely results from an unconventional origin of the CDW and associated charge fluctuations.\cite{reznik2012phonon,raichle2011highly,pintschovius2005electron,PhysRevB.73.224502,reznik2008photoemission,park2014evidence}

We found that DFT calculations of phonons in Hg1201 give different phonon energies depending on the approximations taken and, consequently, that their predictive power is insufficient. The overall good agreement between the data and DFT illustrated by Fig. 1 was obtained by trying different approximations until the best agreement was obtained. Based on this result, we conjecture that other methods of calculating lattice dynamics for this system are equally ambiguous and that it is essential to perform experiments to obtain the correct picture of the lattice dynamics. Our calculation as well as that of ref. \citenum{bauer2010phonon} give good agreement of the overall softening of bond-stretching phonons at the zone boundary for the highly underdoped sample. For other cuprates DFT calculations reproduce the dispersions in strongly overdoped samples\cite{reznik2008photoemission,PhysRevB.73.224502}. It has been suggested\cite{bauer2010phonon,Bauer_2010_2} that failure of the DFT to reproduce the dispersion dip in superconducting samples may be due to charge inhomogeneities not taken into account in the calculation of the phonon dispersion. This point should be addressed in future studies by looking to intermediate doping.

Ref. \citenum{uchiyama2004softening} reported anomalous softening and broadening of Cu-O bond-stretching phonons half-way to the zone boundary along [100], similar to phonon anomalies in other cuprates. However, only one peak was clearly seen in the IXS raw data whereas lattice dynamical calculations give a complicated picture of phonons at energies near the bond-stretching phonon (60-80 meV), with four branches possibly contributing to the signal \cite{bauer2010phonon,Bauer_2010_2}. Thus it is possible that some, or all of the anomalous behavior reported in Ref. \citenum{uchiyama2004softening} originates from interactions among these branches, and not from electron-phonon coupling. The analysis in Ref.  \citenum{uchiyama2004softening} followed the standard procedure to assume that the background underneath the bond-stretching phonon peak is entirely due to resolution function tails of the elastic line and low-energy phonons, i.e. it was not adjustable. 
Ref. \citenum{Bauer_2010_2} showed that the IXS data of Ref. \citenum{uchiyama2004softening} could be fit to four peaks if the background is adjusted to be lower. Just like the one-peak fit, the four-peak fit also reveals a larger bond-stretching dispersion anomaly than we observe in our underdoped sample. We attribute these differences primarily to differences in hole doping levels. This result indicates that the phonon anomaly is enhanced at optimal doping compared to the underdoped composition, similar to previous results for LSCO \cite{reznik2006electron,park2014evidence}. This stronger softening at optimal doping may lead to a mixing with the bond-bending branch, which was not observed in the X-ray measurements as a separate peak. \cite{uchiyama2004softening} Future neutron scattering measurements of optimally-doped Hg1201 should allow a more accurate determination of the dispersion of all phonons and of the effects of branch crossings. In addition, it will be illuminating to see if any new phonon anomalies appear at intermediate doping, where the CDW correlations are strongest. \cite{blackburn2015x,comin2016resonant,Ghiringhelli821}

We also determined that the silent bond-buckling mode, which could be mistaken for a magnetic excitation \cite{fong1995phonon}, appears at 31 meV. In contrast to YBCO, \cite{reznik1994high,raichle2011highly} its dispersion in Hg1201 is not consistent with significant electron-phonon coupling. 

Neutron scattering measurements of moderately-doped and optimally-doped Hg1201 revealed a dispersionless magnetic mode with energy $\sim 53$ meV throughout the entire Brillouin zone, and the mode was associated with the pseudogap formation. \cite{YuanLi2010,Li2012} We did not observe this mode, presumably because its strength considerably weakens at low doping. We also note that there are no phonons near this energy in the experiment and in our DFT calculations (see Fig. 1f). Consequently, this mode reported at higher doping levels is not associated with phonons. 

To conclude, we find qualitative agreement between experimental and calculated phonon dispersions for a highly underdoped sample of the prototypical copper-oxide superconductor $\text{HgBa}_2\text{CuO}_{4+\delta}$, except for the experimentally-observed phonon anomaly in the bond-stretching branch, which is absent in the calculations. This anomaly appears to be weaker than at optimal doping, where it was previously observed via inelastic X-ray scattering. This result implies that significant low-energy charge fluctuations that renormalize phonons, extend to the strongly underdoped side of the superconducting dome in $\text{HgBa}_2\text{CuO}_{4+\delta}$, and it demonstrates the universality of this phenomenon in the cuprates. We hope that the present work will motivate a new generation of quantitative investigations of phonon spectra in other quantum materials.

\section{acknowledgements}
We thank M. K. Chan, Y. Tang, and G. Yu for their help with sample preparation and R. Heid for helpful discussions and preliminary DFT calculations. The work at the University of Colorado was supported by the DOE, Office of Basic Energy Sciences, Office of Science, under Contract No. DE-SC0006939. The work at the University of Minnesota was funded by the Department of Energy through the University of Minnesota Center for Quantum Materials under DE-SC-0016371. This research used resources at the Spallation Neutron Source, a DOE Office of Science User Facility operated by the Oak Ridge National Laboratory.

\bibliography{HgBa2CuO4}

\end{document}